# Precoding in Multibeam Satellite Communications: Present and Future Challenges

*Miguel Ángel Vázquez[1], Ana Pérez-Neira[1,2](IEEE Senior Member), Dimitrios Christopoulos[3](IEEE Member), Symeon Chatzinotas[3](IEEE Senior Member), Björn Ottersten[3] (IEEE Fellow), Pantelis-Daniel Arapoglou[4], Alberto Ginesi[4], Giorgio Taricco[5](IEEE Fellow).*

**Abstract**

Whenever multibeam satellite systems target very aggressive frequency reuse in their coverage area, inter-beam interference becomes the major obstacle for increasing the overall system throughput. As a matter of fact, users located at the beam edges suffer from a very large interference for even a moderately aggressive planning of reuse-2. Although solutions for inter-beam interference management have been investigated at the satellite terminal, it turns out that the performance improvement does not justify the increased terminal complexity and cost. In this article, we pay attention to interference mitigation techniques that take place at the transmitter (i.e. the gateway). Based on this understanding, we provide our vision on advanced precoding techniques and user clustering methods for multibeam broadband fixed satellite communications. We also discuss practical challenges to deploy precoding schemes and the support introduced in the recently published DVB-S2X standard. Future challenges for novel configurations employing precoding are also provided.

## Introduction

Mimicking the terrestrial cellular networks, satellite systems providing broadband IP services are moving from single beam to multibeam architectures, typically operating in Ka-band. In this context, the payload is equipped with a multiplicity of feeds so that information is simultaneously sent to different spot beams on ground with a certain frequency reuse pattern. In such a configuration, bandwidth can be efficiently reused in beams sufficiently separated.

An end-to-end multibeam system architecture from the gateway (GW) to the user terminals (UTs) is depicted in Figure 1, where it can be observed that UTs receive and transmit information through the multibeam antenna pattern, which is fed by the feeder link. The available bandwidth in the feeder link must be large enough to support the frequency re-use adopted for the user beams.

Miguel Ángel Vázquez is the corresponding author.
[1] Miguel Ángel Vázquez and Ana Pérez-Neira are with the Centre Tecnològic de les Telecommunicacions de Catalunya. {mavazquez,aperez}@cttc.es
[2] Ana Pérez-Neira is with the Universitat Politècnica de Catalunya. ana.isabel.perez@upc.edu.
[3] Dimitrios Christopoulos, Symeon Chatzinotas, Björn Ottersten are with Interdisciplinary Centre for Security, Reliability and Trust (SnT), University of Luxembourg. {dimitrios.christopoulos, symeon.chatzinotas, bjorn.ottersten}@uni.lu
[4] Alberto Ginesi and Pantelis-Daniel Arapoglou are with ESA's Research and Technology Centre (ESTEC). {Alberto.Ginesi, pantelis-daniel.arapoglou}@esa.int
[5] Giorgio Taricco is with the Politecnico di Torino. taricco@polito.it

The projection of the beam generation process on Earth by the satellite allows for different information streams at different rates sent towards each beam separately, combined with adaptive coding and modulation (ACM) to match the underlying channel conditions. However, reusing the frequency over the multiple spot beams generates co-channel interference among adjacent ones. Indeed, as there is no way of having completely isolated spot beams, a carefully planned overlap between them is designed, leading to a controlled carrier to co-channel interference ratio C/I.

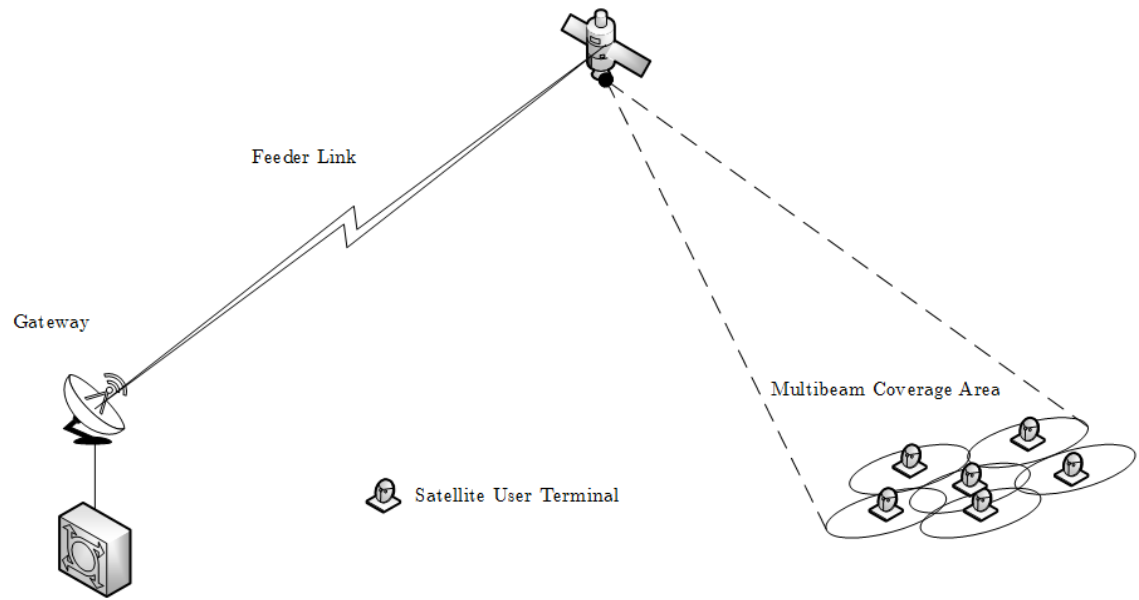

**Figure 1 Multibeam satellite system architecture.**

Consequently, the transmit signal toward a certain beam is partially radiated to the corresponding adjacent beams through its antenna sidelobes. Although the received power levels of the adjacent beams are not as large as the intended one, the created interference accumulates and, therefore, the communication links suffer from a degradation of the achievable signal-to-interference and noise ratio (SINR) that depends on frequency reuse scheme and the UT position.

In order to cope with this intrasystem interference, satellite operators and manufactures try to separate the frequency bands of the adjacent beams so that the interference is mitigated. Commonly for high throughput satellite (HTS) systems, the frequency band is divided into four sub-bands so that closer adjacent beams have disjointed frequency bands (see rightmost part of Figure 2). Another degree of separation between the spot beams is offered by employing orthogonal polarizations.

Although this 'cake-cutting' spectrum management sufficiently alleviates the problem of multibeam interference, it precludes scaling the system capacity by using all the available bandwidth in every beam. Therefore, for achieving a huge leap in offered data rate services, a more aggressive use of the spectrum is desirable. Notably, this vision has been also promoted in cellular system by shifting the paradigm from single cell to multicell systems [2].

As a result, high frequency reuse combined with appropriate interference mitigation techniques is a natural way forward for this scenario. Nevertheless, if the transmitter is devoted to mitigate interference, a certain amount of resources must be dedicated to feedback information from the receiver to the transmitter, which creates additional signaling overhead. Moreover, if the feedback information suffers from a degradation (quantization errors, outdated information, transmission errors,…), the ability to perform interference mitigation is compromised.

Indeed, deciding whether the interference should be removed either at the transmitter or the receiver in satellite systems needs a careful study encompassing all the performance metrics and hardware designs. This paper focuses on the benefits of precoding techniques (i.e. transmitter based interference mitigation) as recent results have demonstrated its large potential yet maintaining the receivers complexity low [5].

Prior to presenting the current multibeam satellite precoding techniques, the channel model is revisited as well as the channel state information reporting method which the UTs perform.

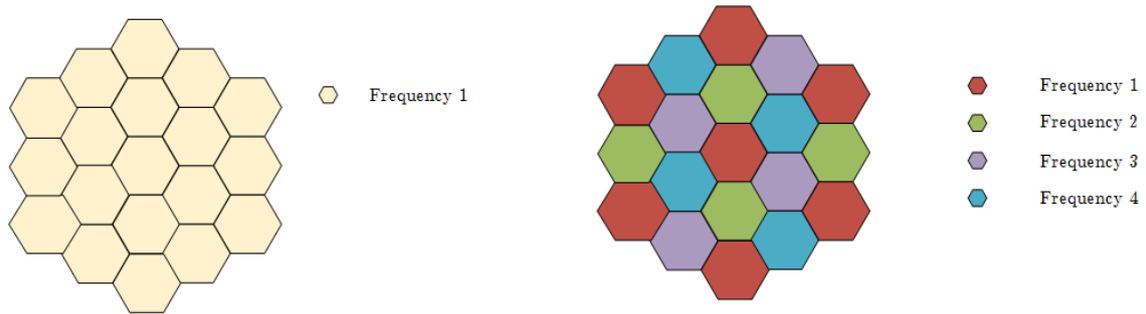

**Figure 2 Frequency reuse patterns in Multibeam satellite communications. Left: full frequency reuse, right four colour-frequency reuse.**

# Channel modelling

In the following, we highlight the components that need to be included in the full channel matrix of a multibeam satellite system in order to be representative of a real system. Since the channel matrix comprises complex elements, particular emphasis is given on the phase effects that need to be accounted for in the channel matrix, as they critically impact the calculation of the precoding matrix. Moreover, accurately estimating channel phase effects is extremely challenging in long haul satellite systems, as the round trip delay from a geostationary satellite corresponds to about 500 ms.

In general, the entire route from transmission to reception (including all the analog & RF circuits, antennas and propagation) should be part of the channel definition, affecting both its amplitude and phase. All other link budget parameters being fixed, the channel amplitude $|h_{kn}|$ between receive antenna $k$ and transmit antenna $n$ of each element in the channel matrix depends on the satellite antenna gain and the propagation effects, predominantly rain attenuation, which is slowly varying and so is $|h_{kn}|$.

Concerning the various contributors that rotate the carrier phase and create a time variant channel phase between receive antenna $k$ and transmit antenna $n$ $\theta_{kn}(t)$ are:

- In the feeder link, the gateway local oscillator (LO) along with its frequency/phase instabilities, as well as the feeder link geometry. However, these phase effects are common for all the beams served by the same gateway.
- In the satellite payload, the LOs of the frequency converters along with their frequency/phase instabilities. Also, the movement of the satellite within its station keeping box modifies both the user and feeder link geometries making the phase time variant. These effects are common to all user terminals in the coverage area.
- In the user link, the user link geometry and the LO of the user terminal's receiver low noise block (LNB). These effects are particular to each user terminal in the coverage area and are removed by the phase synchronization loop.

Assuming an ideal feeder link, Figure 3 provides an overview of the various phase contributions for a simplified scenario involving two payload chains and two user terminal beams on ground.

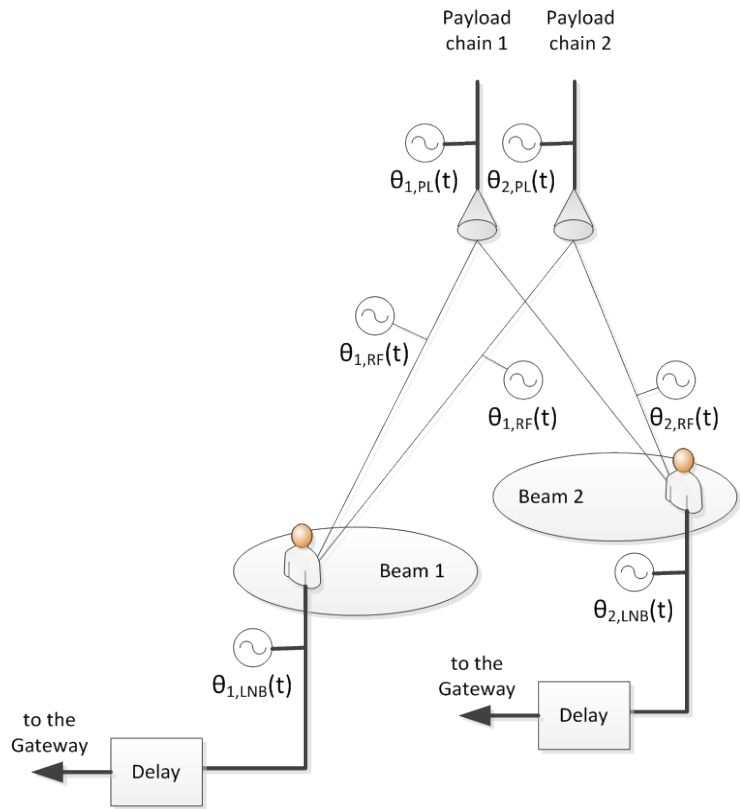

**Figure 3 Phase contributions to the channel matrix.**

In addition, and in contrast to the cellular systems, in fixed satellite systems the channel magnitude does not vary significantly. This is due to the fact that the channel propagation is dominated by the line-of-sight component and slow atmospheric fading. However, the phase variations are not negligible and they constitute the major source of channel impairments as we will discuss in a later section. The phase estimation of any complex sub-channel is as follows:

1) The UT makes a data aided estimate of the incoming signal

$$\hat{\theta}_{tot}(t) = \theta_{tot}(t) + e(t)$$

2) The UT periodically provides phase reports $\hat{\theta}_{tot}(t + \Delta t)$ every $\Delta t$ to the gateway.

3) The gateway will apply this phase estimate (as well as the amplitude estimate) to compute the precoding weights and will transmit the forward link signal toward the UT (assuming zero queuing and processing delay). The signal will be received at the UT after an additional two hop propagation delay of 250 ms. In total, the UT phase estimate $\hat{\theta}_{tot}(t)$ will be outdated by at least 500-600 ms.

Although this channel reporting mechanism is simple, its performance is crucial to the final precoding desing. Indeed, as it is described in the next subsections, the channel state information (CSI) errors reduce the system throughput. Prior to investigating the effects of non perfect CSI, the precoding techniques are presented next.

## Overview of Multibeam Precoding Techniques

The role of precoding is to 'revert' the multibeam interference signals so that the receiver can perform single user detection yet maintaining a high SINR. In this way, the additional spectrum made available through higher frequency re-use can be exploited and a larger system capacity can be delivered compared to the orthogonal transmission one.

As it happens in cellular networks, a condition for efficient forward link precoding is that the receivers provide high quality reports of a number of sub-channels back to the transmitter that is responsible of deriving the appropriate precoding matrix. In the next paragraphs, we first present the design problem as a set of proposed solutions and their implementation in the recent DVB-S2X (digital video broadcasting over satellite 2 extension) standard.

**Multibeam Satellite Precoding Algorithms: Multigroup Multicast Precoding**

Apart from the difficulties in obtaining up-to-date channel state information at the transmitter (CSIT), satellite precoding substantially differs from its terrestrial counterpart due to its multicast operation, which stems from the fact that a single DVB-S2X codeword (frame) contains data bits addressing multiple UTs. Although there is a current tendency to add multicast modes to the cellular standards, the general operation is a simple multiuser model where a certain number of independent symbols, *K*, are intended to be transmitted to *K* users. In contrast, in satellite communications a given codeword always needs to be decoded by more than one receiver. This phenomenon entails a decrease of the offered spectral efficiency since the UT with the lowest SINR determines the rate allocation via ACM. In this context, the multibeam satellite forward link transmission is modelled by the multigroup multicast multiple-input-single-output (MISO) channel.

From early multibeam precoding studies, it was evident that linear algorithms considering only one user per frame were performing satisfactorily providing significant capacity gains [4]. One of the most promising designs, initially conceived for cellular communications, was the regularized channel inversion, also coined minimum square error (MMSE) due to its similarity to the linear estimator. As a result, the first attempt of designing a multicast multigroup precoding was a modification of the regularized channel inversion. This was carried out in [5] where all the users served by the same frame are considered as a single user with an equivalent channel equal to the average of the different channels. Spectral efficiency results are encouraging considering this simple approximation; however, the channel averaging degrades when several users are considered per frame. In order to potentially improve performance, a technique based on the block singular value decomposition (SVD) method as presented in [6] has been investigated. In that case, the precoding matrix is constructed row by row, computing each time the null space with the aim of completely rejecting the interference to the other beams. Although the computational complexity increases severely, a certain gain in terms of spectral efficiency is obtained.

Towards obtaining the sum-rate optimal design, an optimization approach is provided in [7], where the authors formulate the optimization problem of multicast multigroup with per antenna power constraints. Although optimal, this approach suffers from high computational complexity, which should be considered in order to accommodate it in commercial satellite ground segment equipment.

Figure 4 depicts a 100-runs simulation for a Ka-band multibeam satellite system consisting of 245-spot beams. For further details of the channel model and particularly the antenna pattern, the reader can refer to [8], where the overall channel effects are presented. The three precoding techniques briefly introduced above were considered, namely 'MMSE', 'block SVD' and 'Frame-based'. It can be observed that frame-based precoding offers a larger per-beam throughput than the other proposed techniques. Indeed, MMSE precoding shows for this case a very low performance. Indeed, compared to the conventional 4-color case which offers 0.73 Gbits/s average throughput per beam for any number of users, MMSE presents lower average throughput. On the other hand, for the more favorable case of two users per frame, Frame-based and Block-SVD gains approximately 135% and 120% over the reference 4-color case.

As a matter of fact, the larger number of users per frame is targeted, the lower the throughput gain obtained, and the three techniques present a similar decay of throughput with the number of users. The reason for this decrease is the differences between the channel vectors of the users served sequentially by the same precoding matrix. In other words, whenever all channel vectors of the users in the same beam are collinear; there will be no loss of performance due to the multicast transmission. Selecting which users to group together and apply precoding on is the task of a scheduler, which will be discussed in the next section.

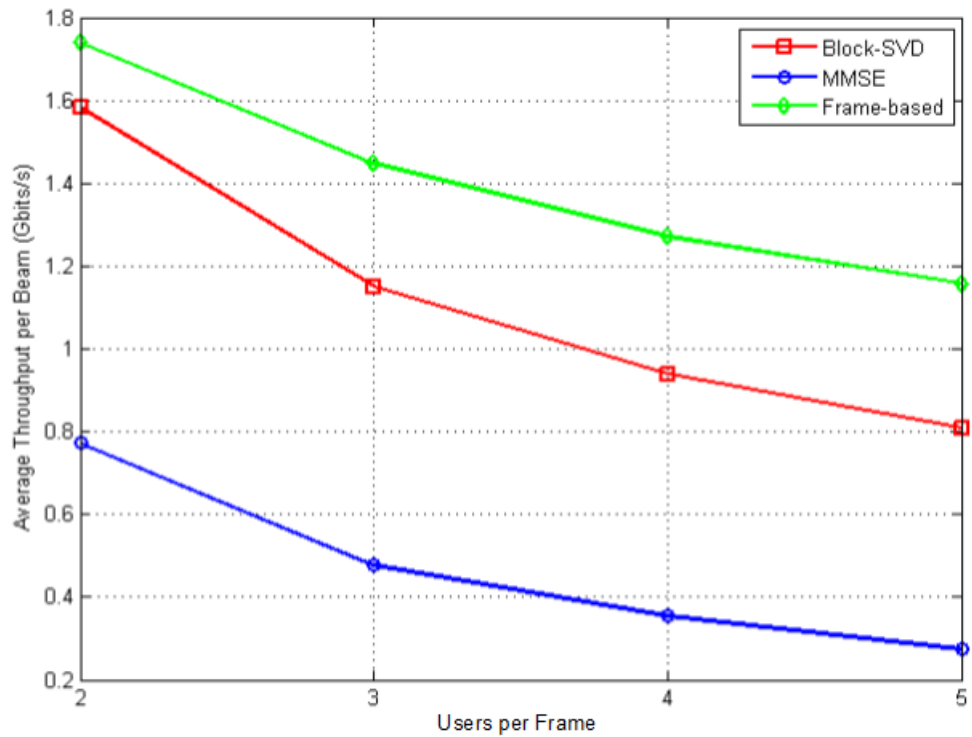

Figure 4 Average throughput per beam versus number of users per beam. The throughput obtained via 4-color frequency reuse pattern and without using precoding is 0.73 Gbits per beam. Remarkably, for more than 3 users per beam, the reference 4-color frequency reuse offers higher throughput than the MMSE precoding. Frame-based precoding presents the highest throughput for this scenario.

**User Scheduling and Precoding**

Addressing more than one user in the same frame increases the precoding complexity as different spatial signatures need to be considered for a single codeword. The different users belonging to the same beam will have different channel vectors, leading to poor precoding performance. However, the system designer can smartly choose the users so that they have *similar* channel vectors. Under this context, the scheduler must not only consider the SINR levels for grouping the users as it does for DVB-S2, but also its channel vector.

Finding out which are the best users to be served in a certain time instant considering that they are going to be precoded with the same precoding weights is a very challenging problem. The reason is that after the precoding effect, the SINR levels for each user vary so that the a priori modulation and coding (modcod) assignment becomes obsolete. Therefore, the optimization is a computational demanding iterative process as described in [9].

Whenever a low complexity non-iterative process is targeted, certain heuristics must be determined in order to come up with an easy-to-implement user grouping. A pragmatic approach in this direction can be found in [5], where a geographical user clustering (GUC) is proposed. This technique refers to the user grouping carried out taking only into account the channel matrices without the phase effect. According to GUC, UTs closely located are clustered in the same group so that their information will be encoded in the same codeword (frame). The implementation of this technique can be performed without large computational complexity. Given a certain set of users to be served with their respective channel vectors, the *k*-means algorithm groups the users with the similar channel vector.

Assuming that the GW has access to both channel amplitude and phase effects, a more accurate clustering can be done. It is evident that the phase effects increase the differences between the channel vectors of the users in the same beam, leading to a drastic reduction of the overall performance.

Yet a better heuristic user clustering can be conceived whenever we assume that the GW has full CSIT. This method, coined as frame based clustering (FBC), is described in [10]. The key step for the FBC lies in measuring the similarity between user channels based on the readily available CSI. The underlying intuition is that users scheduled in the same frame should have co-linear (i.e. similar) channels, while interfering users, scheduled in adjacent synchronous frames, should be orthogonal to minimize interferences.

Figure 5 presents a simulation where the different precoding and user clustering methods are presented. Perfect clustering refers to the technique of user grouping considering not only the magnitude of each channel but also the phase. On the contrary, GUC refers to the technique which only takes into account the magnitude of the channel (i.e. its position). Finally, frame-based clustering considers an heuristic approach described in [8]. Consider the Block-SVD and MMSE methods; it can be observed that whenever the scheduler uses all the available CSIT (i.e. phase and magnitude), larger throughputs can be obtained. Note that even the combination of scheduling and precoding with the lowest performance (i.e. MMSE with GUC), is better than not taking into consideration the user clustering problem (see Figure 4 for comparison). Consequently, user clustering is an important tool for applying precoding techniques in next generation multibeam broadband satellite systems. It is worth remarking that the combination of scheduling and precoding that offers the largest performance is Block-SVD jointly with perfect clustering. Finally, it is important all methods offer larger throughputs than without considering the user clustering (Figure 4).

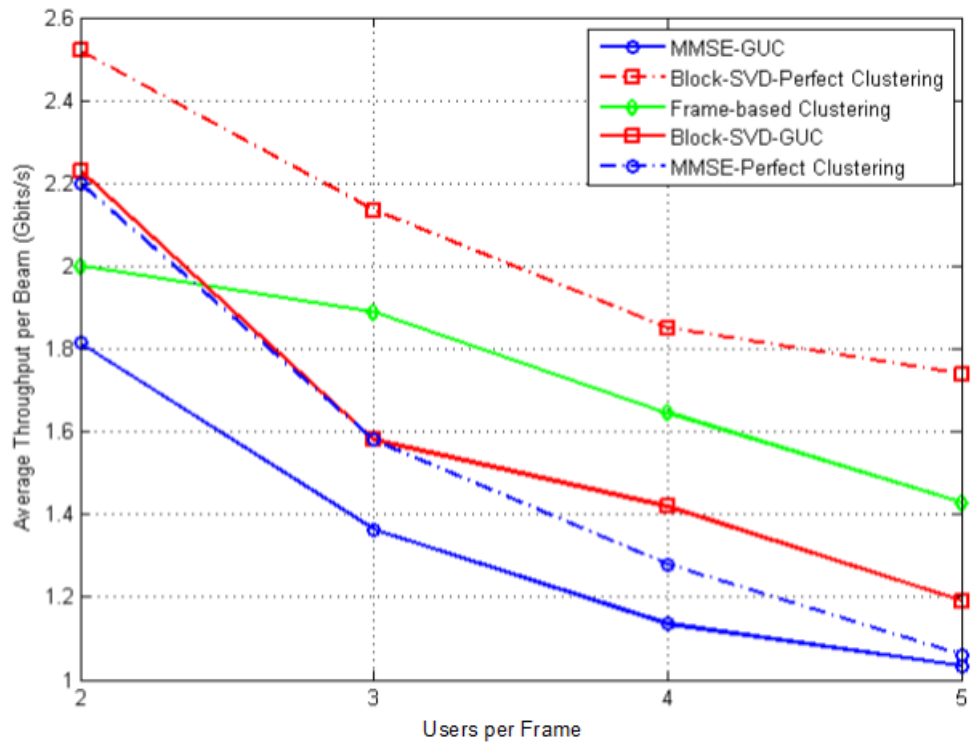

**Figure 5** Average throughput per beam versus users per beam when considering different user clustering techniques. Perfect clustering refers to the technique of user grouping considering not only the magnitude of the channel but also the phase effect. On the contrary, GUC refers to the technique which only takes into account the magnitude of the channel (i.e. its position). Finally, frame-based clustering considers a heuristic approach described in [8].

Apart from the receiving the corresponding CSI information, the enabling communication protocol must provide certain synchronization and channel estimation aspects. This is described in the following subsection and, posteriorly, the performance decrease when the CSI is not accurate at the gateway is also analyzed.

**DVB-S2X specifications for precoding**

The potential of precoding in a broadband multibeam satellite system is also linked to the support for this technique included in relevant standards. In terms of standardization developments, the DVB very recently (2014) issued the DVB-S2 extension (DVB-S2X) with an optional specification that provides the necessary framing and signalling support to interference management techniques (Annex E of [3]). Thereby, all signalling and framing elements to support precoding in terms of physical layer have been put in place. The specification in [3] endorses a number of special features including regular channel framing structure, specific pilots and unique words for synchronization aid as well as a feedback signalling message from the UTs to the GW.

The so-called super-frame structure supports orthogonal Start of Super-Frame (SOSF) and pilot fields by using Walsh-Hadamard sequences. A set of orthogonal sequences can be assigned to co-channel carriers within a multi-spot beam system (a unique sequence per beam). These features allow the UT to estimate the channel responses down to a very low SNIR value. The beam-specific orthogonal sequence allows the receiver to uniquely associate the channel estimate to the beam index. The superframe also foresees an additional precoded pilot field to help amplitude and phase recovery in support of the precoded data detection. Apart from the aforementioned features, the use of this super-frame extensively helps the use of precoding due to its multicast transmission nature.

Indeed, in DVB-S2 as well as in DVB-S2X, more than one user are allocated in the same frame in order to allow for large LDPC (low density parity check) codewords, which increases the channel coding gain and the frame encapsulation efficiency. Under this context, the precoding matrix needs to be computed each time there is a change on the FEC block (i.e. group of users' coded information) for all beams. This is shown in the upper part Figure 6 where the dotted vertical line indicates each time the precoder would need to be re-computed. Aside of the computational complexity of computing the precoding matrix that often, the frame format in the upper part of Figure 6 does not allow for the alignment of pilot fields among the beams. Nevertheless, it should be pointed out that the more often the precoding matrix is computed, the larger spectral efficiencies can be obtained.

In order to solve this problem, DVB-S2X introduced the super-frame structure shown in the lower part of Figure 6, with the same number of symbols so that precoding can be applied in a per-super-frame basis. This structure aligns the pilot fields among the beams and also yields a periodic precoding matrix refresh period. Moreover, the fact that the frames are time aligned, leads to a substantial reduction of the receiver complexity.

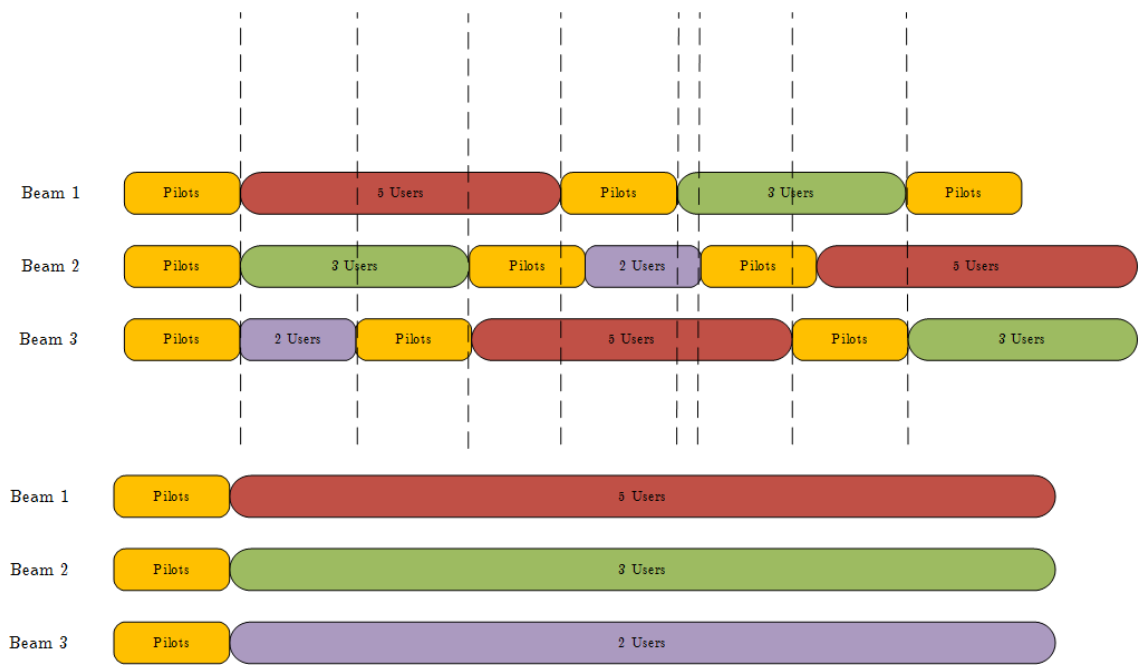

**Figure 6 Frame transmission in DVB-S2 (up) and DVB-S2X (down). The use of precoding depends on the spatial signature of the receivers so that in the DVB-S2 structure, the precoding matrix need to be updated at each dotted line.**

Focusing on the CSI feedback, when a super-frame is transmitted each UT estimates a number of channel complex coefficients corresponding to the most significant interfering beams (16 or 32). These complex numbers are quantized and signalled back to the gateway using the signalling message described in Annex E of [3] with a maximum rate of one message every 500 ms.

**Prospective Gains under channel CSIT error**

Although the results presented hitherto are very promising, they do not account for practical impairments experienced in a real system. In order to determine the precoding performance under CSIT imperfections, a simulation is performed considering that the transmitter receives a perfect estimation of the channel magnitude whereas the channel phase suffers from certain deviation. Remarkably, this is a close-to-real simulation set up since the channel magnitude can be easily known at the GW with negligible error as it exhibits slow dynamics. For this case, the phase uncertainty is modelled by an additional Gaussian random phase with zero mean and 10º variance (which is a rather pessimistic assumption). Figure 7 shows the dramatic decrease of performance with respect to the perfect CSI case depicted in Figure 5 when considering Block-SVD-GUC precoding. On the other hand, MMSE-GUC is more robust to CSI errors. Indeed, although MMSE presents certain throughput decrease with respect to the perfect CSIT case, Block-SVD precoding is more sensitive to these imperfections.

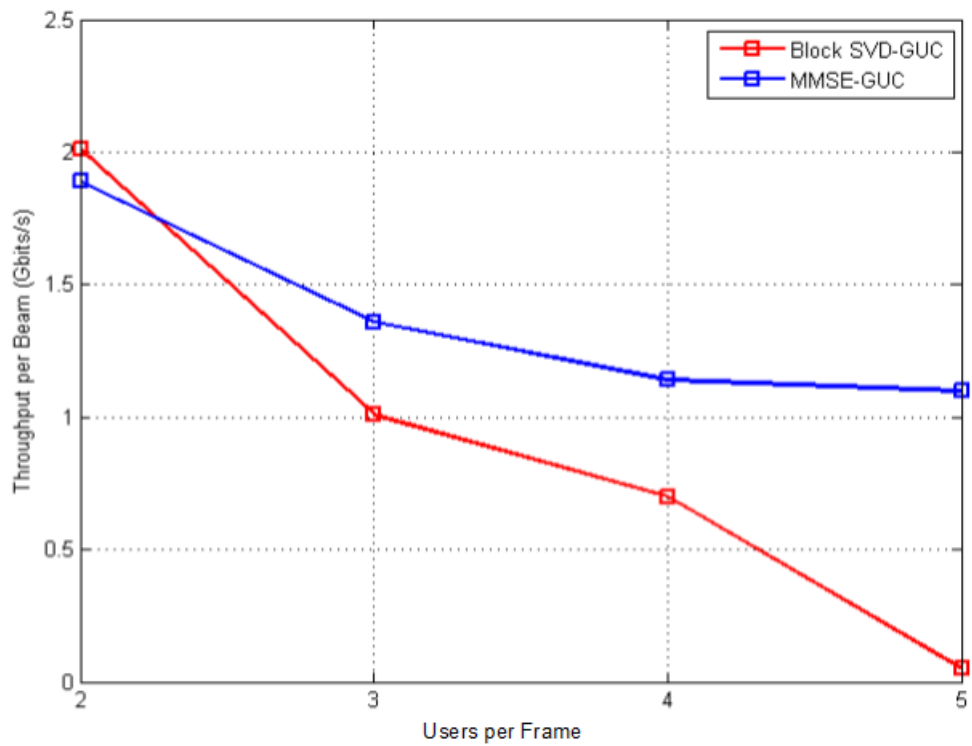

**Figure 7 Average throughput per beam versus number of users considering imperfect CSIT. MMSE with is more robust to channel uncertainties than Block-SVD as it does not present such a performance decrease with respect to the perfect CSI case (Figure 5). Note that when more than 4 users are considered, the Block-SVD approach presents lower throughput than the 4-color scenario.**

# Future Trends: Multi-Gateway Configurations in Precoding

Apart from the presented issues on precoding in multibeam satellite systems, other challenges are appearing. One of the most important ones is feeder link bottleneck bandwidth and how to prevent it. Indeed, a common assumption in the literature on multibeam precoding has been ideal and unlimited bandwidth feeder link. Nonetheless, the consideration that all spot beams in the system are served by a single GW is not practical purely due to the limited feeder link spectrum even in a conventional four colour system, when a large number (>200) of beams is used. The situation is of course aggravated by moving to more aggressive frequency re-use architectures. Although there are efforts to move the feeder link to higher frequency bands like Q/V-band (40/50 GHz) and W-band (70/80 GHz) [11] to moderate the number of GWs, for large systems a multiplicity of GWs should be accounted for.

The effects of multiple GWs in the precoding design have been studied in [12] leading to a performance loss with respect to single GW precoding due to the fact that there is no longer a single transmitter entity in possession of the CSIT for all UTs. This allows the GWs to pre-compensate for the co-channel interference of the UTs only for the subset of beams that each is serving (cluster). Notwithstanding this performance loss, noticeable gains by GW cooperation under the beam clustering assumptions are still reported in [12], where the deployment of non-cooperative multiple GWs improves the system performance over the conventional four color frequency reuse systems by almost 30%. The limiting factor in this case is the interference from adjacent beams that belong to different GWs, particularly in cluster edges. If no cooperation exists between the GWs joint processing cannot mitigate interferences from beams served by different GWs. To this end, partial cooperation amongst GWs that serve adjacent beam-clusters is proposed.

# Conclusions

This paper provides a high level overview of advanced precoding techniques, which we strongly believe will become the key technology for the next generation of very high throughput systems employing multibeam satellites. High level requirements and practical aspects related to this technology, such as channel impairments and user grouping, as well as the DVB-S2X standard adaptation for precoding, are discussed. A performance evaluation based on the current DVB-S2X standard quantified the benefit of this technique. This gain demonstrates that if multibeam satellite networks incorporate precoding mechanisms, they will be able to face the challenge of ever increasing broadband Internet traffic.